# Real-time diffuse correlation spectroscopy with a chip-based correlator for measuring human cerebral blood flow and brain function

**Quan Wang,[1,2,4,5] Yuanyuan Hua,[3] Chenxu Li,[1] Zhizheng Yuan,[2] Jing Wang,[2] Ahmet T. Erdogan,[3] Hanqing Chen,[2] Xunting Huang,[2] Maciej Wojtkiewicz,[3] Alistair Gorman,[3] Mingliang Pan,[1] Yuanzhe Zhang,[1] Yining Wang,[3] Neil Finlayson,[3] Renzhe Bi,[6] Robert K. Henderson,[3] Zhen Yuan,[2,4,5*] and David Day-Uei Li[1,*]**

[1]*University of Strathclyde, Faculty of Engineering, Department of Biomedical Engineering, Glasgow, UK*

[2]*Faculty of Health Sciences, University of Macau, Macau SAR, China*

[3]*The University of Edinburgh, School of Engineering, Integrated Nano and Micro Systems (IMNS), Edinburgh, EH9 3JL, UK*

[4]*Zhuhai UM Science and Technology Research Institute, University of Macau, Hengqin, China*

[5]*Centre for Cognitive and Brain Sciences, University of Macau, Macau SAR, China*

[6]*A*STAR Skin Research Labs (A*SRL), Agency for Science, Technology and Research (A*STAR), 31 Biopolis Way, Nanos, Singapore 138669, Singapore*

*david.li@strath.ac.uk and zhenyuan@um.edu.mo

**Abstract:** Diffuse correlation spectroscopy (DCS) is a non-invasive optical technique that probes microvascular blood flow in deep tissues. Here, we present and validate a new on-chip hardware correlator for high-speed DCS measurements. The correlator is embedded in a custom-built 512 × 512 single-photon avalanche diode (SPAD) array named ATLAS, which computes intensity autocorrelation functions directly on-chip at a sampling rate of 116 Hz – the fastest DCS acquisition reported to date. Unlike conventional DCS systems that suffer from low light throughput and therefore cannot resolve cardiac pulsations at source-detector separations ($\rho$) beyond 30 mm, our massively parallel on-chip architecture computes autocorrelations within each macropixel, eliminating the data-throughput bottleneck. This enables high-SNR, real-time detection of pulsatile blood flow even at $\rho$ = 50 mm on the human forehead. In phantom experiments at $\rho$ = 25 mm, ATLAS-DCS achieves a 12-fold improvement in signal-to-noise ratio over a conventional single-channel DCS instrument while operating at 116 Hz. In human subjects, we resolve functional hyperemia during a mental arithmetic task at $\rho$ = 30 mm. Furthermore, we integrate ATLAS-DCS with a frequency-domain near-infrared spectroscopy (FD-NIRS) module, enabling simultaneous monitoring of blood flow and tissue oxygenation. With this combined system, we can concurrently resolve core hemodynamic parameters. The on-chip parallelized DCS design substantially improves detection speed, depth sensitivity, and real-time capability, paving the way for wearable, high-speed cerebral blood flow monitoring in both clinical and research settings.

## 1. Introduction

Cerebral blood flow (CBF) is a critical indicator of brain health, as it indicates whether the brain receives sufficient oxygen and nutrients to maintain its functions [1]. Abnormal CBF can lead to severe neurological disorders, including ischemic stroke [2,3], traumatic brain injury [4,5], neurodegenerative diseases [6], Alzheimer's disease [7,8], and Parkinson's diseases [9,10]. In addition, CBF also reveals brain functionality: neural activity triggers hemodynamic changes via neurovascular coupling [11–13], meaning that neuronal activation induces localized blood flow (BF) variations. Continuous, real-time monitoring of CBF is therefore highly desirable to enable timely intervention and personalized treatment.

Diffuse correlation spectroscopy (DCS) is an optical technique that can measure human CBF based on coherent light re emitted from tissue [14–17]. The resulting BF index (BFi), a parameter directly proportional to actual BF is obtained by fitting the autocorrelation function of the speckle intensity to DCS models [18]. This method enables non-invasive, continuous bedside CBF monitoring, which is not possible with other techniques such as positron emission tomography [19] or arterial spin labeling magnetic resonance imaging (MRI) [20]. In DCS, a common rule-of-thumb is that the measurement samples a banana shaped region underneath the optical probe spanning approximately one third to one half of the source detector separation ($\rho$) [21]. Consequently, conventional DCS using $\rho = 20 \sim 30$ mm is confined to superficial cortical regions, corresponding to an estimated penetration depth around 10 ~ 15 mm, leaving deeper cerebral perfusion unmeasured. Although increasing $\rho$ is necessary to achieve deeper penetration, this is hindered by DCS's reliance on single-mode fibers for autocorrelation detection. The inherently low optical throughput of such fibers leads to poor signal -to-noise ratio (SNR) — a critical issue in adults, where $\rho$ must be larger than 20 mm in order to detect cerebral signals. This SNR limitation is further exacerbated by the presence of hair or dark skin, often resulting in a prolonged acquisition time and a degraded temporal resolution [22]. Similar limitations in optical throughput and sensitivity have also been widely observed in fiber-based sensing systems, where advanced photonic crystal fiber and micro-structured fiber designs have been proposed to enhance light–matter interaction and improve detection efficiency [23–26]

A common strategy to mitigate low SNR is to place multiple detector fibers at the same position and average the resulting autocorrelation curves [27]. The drawback is the high cost due to the need for multiple photon counting detectors. Another approach to improve SNR is the use of highly integrated CMOS single photon avalanche diode (SPAD) arrays with multimode fibers or light guides [28], a development that has gained traction in recent years. This trend aligns with broader developments in photonic sensing technologies, where improved optical confinement, birefringence control, and nonlinear enhancement in fiber-based platforms have enabled higher sensitivity and robustness in detection systems [23,25,26,29]. Johansson *et al.* first used a $5 \times 5$ SPAD array to achieve an improved SNR in milk phantoms and *in vivo* blood perfusion tests [30]. This was followed by a $32 \times 32$ SPAD array (Photon Force, Ltd.) [31,32], provided a 32-fold SNR increase. In 2023, Michael *et al.* employed a 500 × 500 SPAD array in optical phantom studies, achieving an SNR gain of nearly 500 [33].

Another fundamental limitation of DCS is the sampling rate, which directly determines the achievable temporal resolution and effective acquisition rate. Conventional DCS BF measurements are typically slow, with a sampling rate around 0.3 ~ 1 Hz. As a result, they can only resolve slow variations over extended time scales, such as resting-state hemodynamics (where BF changes over seconds to minutes). However, many physiologically relevant dynamics demand higher temporal fidelity. For example, evoked responses (e.g., functional activation or cuff occlusion) can induce BF changes within hundreds of milliseconds, requiring a sampling rate >10 Hz to accurately capture the rising and falling edges. Moreover, transient events such as cardiac pulsation, respiration, and motion artifacts occur on a faster time scale. Although the fundamental heartbeat frequency is approximately 1 Hz, the pulsatile components of BF velocity often contain higher-frequency information around 5~10 Hz; thus, a sampling rate of at least 20 Hz is necessary to avoid aliasing.

Several technical advances have been reported to overcome this limitation. In parallel, advances in sensor interface electronics and signal processing strategies have also played an important role in improving measurement robustness and stability in optical sensing systems [34–36]. In 2016, Wang *et al.* [37] achieved a 50–100 Hz sampling rate for DCS using a real-time software correlator based on a "shift-and-add" method. This approach reduced the number of delay channels (e.g., 40 delays within 1–250 μs) and optimized PCIe based photon counting, enabling an integration time as short as 10–20 ms. In 2021, Liu *et al.* demonstrated a 33 Hz sampling rate by parallelizing 1024 SPAD detectors, reducing the autocorrelation integration time to 30 ms [31]. In 2024, Moore *et al.* presented a fully integrated FPGA-based DCS system, achieving a 50 Hz measurement rate by performing autocorrelation and curve fitting on-chip, thereby enabling real-time pulsatile BF monitoring [38]. In 2025, Kreiss *et al.* [39] demonstrated massively parallelized DCS using a 500 × 500 SPAD array, achieving deep-tissue (up to 40 mm source-detector separation) pulsatile BF measurements in human adults at a sampling rates of 8–10 Hz, while maintaining a noise level comparable to a 32×32 array at $\rho$ of 15 mm.

This paper presents the development of a novel hardware accelerated DCS system based on a shift register autocorrelation architecture implemented in custom 3D stacked SPAD technology. Our approach directly computes the intensity autocorrelation function by progressively delaying photon counts through multi-tap shift registers, enabling efficient multiply-accumulate (MAC) operations without streaming raw photon frames. In the custom SPAD array (ATLAS [40]), each macropixel contains a 31-tap, 5-bit shift register and a shared MAC unit operating at a pixel clock of 30 MHz, delivering on-chip autocorrelation with a minimum lag time of 1 μs and an ensemble DCS frame rate of 27,000 fps. Leveraging this shift-register-based data compression and massively parallel architecture (up to 128 × 128 macropixels), we demonstrate real-time measurement of decorrelation time constants from tissue-like phantoms and human foreheads, resolving pulsatile BFat $\rho$ ~ 50 mm. To the best of our knowledge, the experiments represent the highest level of integration reported for DCS to date, combining on-chip autocorrelation computation, high photon detection efficiency (47% at 785 nm), and low power consumption (300 mW). The fast data streams enable accurate tracking of cuff-occlusion-induced hyperemia and cardiac signals, paving the way for wearable, real-time CBF monitoring systems. Additionally, modern data-driven signal processing frameworks have further enhanced the interpretability and robustness of complex measurement signals in biomedical sensing applications [41,42].

The remainder of this paper is organized as follows. We first compare our ATLAS-DCS system with a laboratory-standard DCS system using a tissue-mimicking phantom to validate its measurement accuracy. Next, we validate the system through *in vivo* measurements on the human forearm and forehead. Finally, we present human brain function measurements during a mental subtraction task at $\rho$ = 30 mm, demonstrating clear BF changes that confirm our DCS system's ability to continuously monitor human CBF and brain function with unprecedented signal quality at a $\rho$. This work leverages the on-chip SPAD DCS platform for future applications in biology, cognitive neuroscience, and clinical practice.

## 2. Materials and Methods

### 2.1. *SPAD System*

ATLAS was fabricated using 65/40 nm 3D-stacked CMOS technology, featuring a top-layer with a 512 × 512 array of deep trench isolated microlens SPADs with a 10.17 μm pitch [43]. The SPADs achieve a peak photon detection efficiency (PDE) of 55% at 600 nm, and 26% at 940 nm, with a median dark count rate (DCR) of 500 counts per second at the room temperature under a 23 V bias (the breakdown voltage = 17.8 V). ATLAS comprises a photosensitive array containing 128 × 128 macropixels and a column-parallel processing block. Each macropixel is capable of computing the photon counts $C\tau$ and accumulated photon counts $A\tau$ coefficients at 31 different time lags, utilizing an in-pixel shift register, multiplier, accumulator, and 32 banks of static random-access memory (SRAM). Each macropixel detects incident photons via 16 SPADs, whose outputs are combined and connected to a digital counter. The photon count at each time lag is stored in a 32-element shift register. An in-pixel controller circuit manages the multiplier to calculate the $C\tau$ and $A\tau$ coefficients for the respective time lags, corresponding to the different shift register elements. These coefficients are stored in the in-pixel SRAM memory banks and then read out to the on-chip column-parallel processing circuits, where the $g_2$ coefficient, averaged across the entire macropixel array, is calculated. ATLAS uses two clocks provided by the FPGA: SYS_CLK, which serves as the main clock for the sensor, and PIX_CLK, which manages the macropixels' operations. Both clocks are derived using the mixed mode clock manager. Additionally, a few control signals are required to operate the sensor. The row address bits, ROW_ADDR<0:6>, and channel address bits, CHANNEL_ADDR<0:4> are used to select the desired row of macropixels and the corresponding bank of the in-pixel SRAM memory, respectively. The signals START_ACCUMULATE, START_DIVISION and G2_SUM_RST control the calculations of the $g_2$ coefficient within the column-parallel processing circuits. The $g_2$ data is first read from ATLAS into a first-in-first-out (FIFO) block implemented in the FPGA and then stored in the DRAM. From there, it is transferred to the USB 3.0 microcontroller via another FIFO block, and finally, the data is sent to the PC through the USB 3.0 interface. Fig. 1 shows the configuration of the sensor.

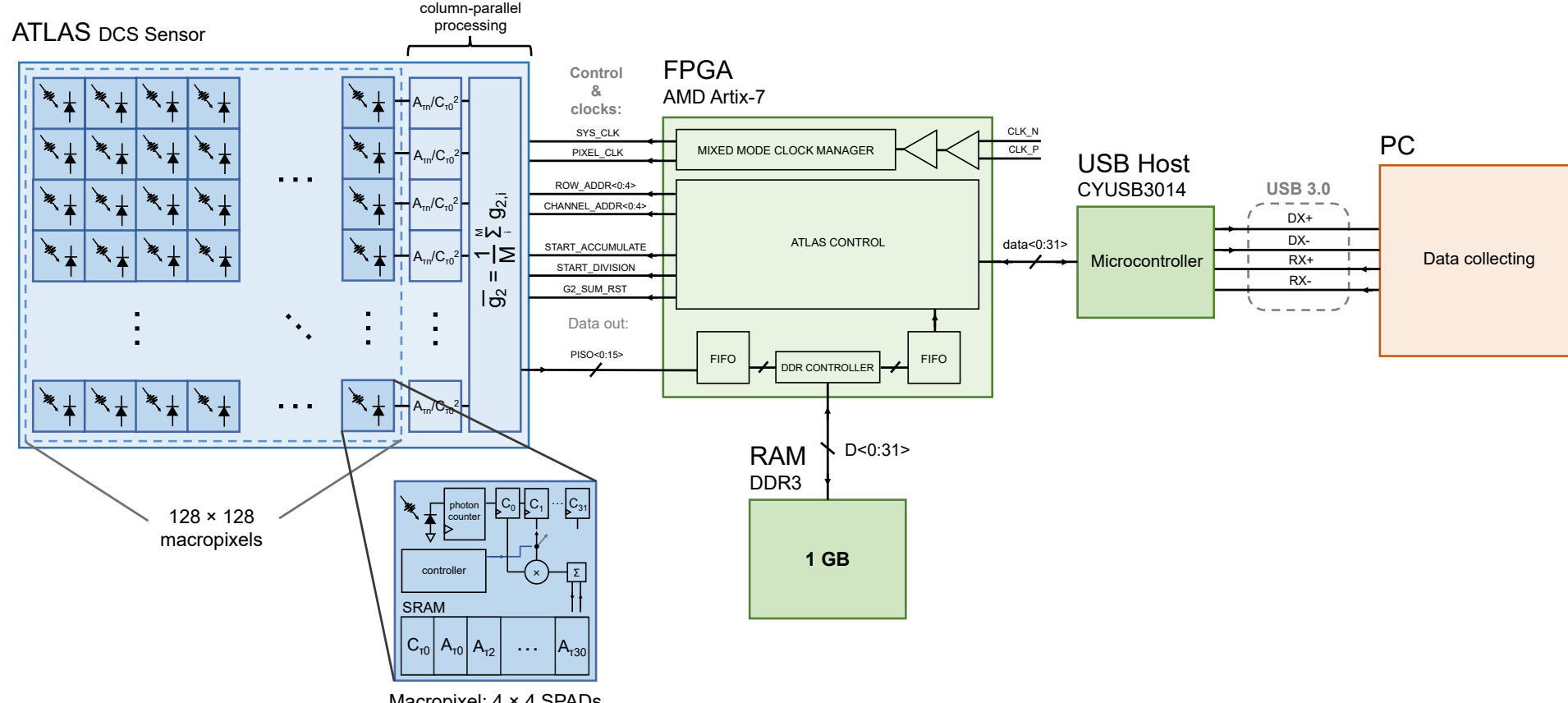

Fig. 1. The DCS camera (using ATLAS) system architecture. ATLAS is controlled and read out by the field programmable gate array (FPGA), employing double data rate (DDR) 3 dynamic random-access memory (DRAM) for fast data acquisition and the USB 3.0 interface to communicate with a PC for data acquisition.

### 2.2. *Theoretical Modelling*

In DCS, the motion information of particles in the tissues is carried in the electric field of diffuse light $E(t)$. It can be extracted from the unnormalized electric field temporal auto correlation function, which is defined as $G_1(\tau) \equiv \langle E(t)E^*(t+\tau)\rangle$, where $\langle\ldots\rangle$ denotes temporal average. $G_1(\tau)$ can be modelled with the correlation diffusion equation [16], giving a normalized solution, $g_1(\tau) = G_1(\tau)/G_1(\tau = 0)$, for a semi-infinite homogeneous geometry under the assumption of extrapolated boundary conditions [44] as,

$$g_1(\tau) = \frac{r_2 \exp\left(-K_D(\tau)r_1 - r_1 \exp\left(-K_D(\tau)r_2\right)\right)}{r_2 \exp\left(-K_D(0)r_1 - r_1 \exp\left(-K_D(0)r_2\right)\right)}, \quad (1)$$

where ${K_D}^2(\tau) = 3\mu_a\mu_s' + \alpha\mu_s'^2 k_0^2 \langle\Delta r^2(\tau)\rangle$, $\mu_a$ and $\mu_s'$ are the absorption and reduced scattering coefficients, respectively. $\rho$ is the distance between the source and detection fibers, α is the probability of photon scattering events from a moving scatterer, $k_0 = 2\pi n_0/\lambda$ is the wavenumber at the wavelength λ, where $n_0$ is the tissue refractive index. And $r_1 = (\rho^2 + {z_0}^2)^{1/2}$, $r_2 = (\rho^2 + (z_0 + 2z_b)^2)^{1/2}$, $z_0 = (\mu_a + \mu_s')^{-1}$ , and $z_b = \frac{2}{3\mu_s'}\frac{1+R_{eff}}{1-R_{eff}}$ with $R_{eff} = 1.4400n^{-2} + 0.7100n^{-1} + 0.6680 + 0.0636n$ being the effective reflection coefficient, which is determined by the ratio of the refraction indices of two media (e.g., $n = \frac{n_0}{n_{air}} \approx 1.37$, $n_{air}$ is the air refraction index). In practice, the mean square displacement, $\langle\Delta r^2(\tau)\rangle$, of the scattering particles in the lags time (τ) can be approximated by the Brownian motion model that shows better fitting in most applications, ranging from muscles to the brain. Thus $\langle\Delta r^2(\tau)\rangle = 6D_b\tau$, where $D_b$ is the effective diffusion coefficient and the product $\alpha D_b$ commonly used as the DCS BFi.

To perform a DCS measurement, $g_2(\tau)$ was calculated at each macropixel SPAD pixel:

$$g_2^j(\tau) = \frac{\langle I_j(t) \cdot I_j(t+\tau) \rangle}{\langle I_j(t) \rangle^2}, \quad (2)$$

where $I_j(t)$ is the number of detected photons of the *j-th* SPAD pixel at a given time *t*. Then the final system autocorrelation function was calculated across $j$ = 1 to *N*; *N* is the number of SPAD pixels used for the measurements,

$$\bar{g}_2(\tau)|_N = \frac{1}{N} \sum_{j=1}^{N} g_2^j(\tau), \quad (3)$$

The normalized intensity ACF, $g_2(\tau)$ is linked to the normalized electric field ACF, $g_1(\tau)$ through the Siegert relation [45]:

$$g_2(\tau) = 1 + \beta |g_1(\tau)|^2, \quad (4)$$

where $\beta$ is the coherence parameter.

### 2.3. *The Experimental Setup*

In our system, a continuous-wave laser (MSL-III-785L, Changchun New Industries Optoelectronics Technology Co., Ltd., Changchun, China) with a wavelength 785 nm was coupled into a multimode optical fiber (HPCF, 400/430/730 µm core/cladding/coating diameter, SC Intelligent Fiber Technology Co., Ltd., Shen Zhen, China) with a numerical aperture (NA) of 0.22. The laser's coherence length (> 2 m) was much longer than typical photon pathlengths, and its maximum output power was 100.2 mW. To attenuate the laser beam output to 30 mW, meeting ANSI safety requirements, a variable neutral density filter (NE205B, OD = 0.5, Thorlabs, UK) was placed between the laser and the source fiber tip.

For conventional DCS, the backscattered light is collected by a single-mode optical fibers with NA of 0.13 (780HP, 5/125/245 µm core/cladding/Coating diameter, SC Intelligent Fiber Technology, Co., Ltd., Shenzhen, China) and sent to a single-photon avalanche diodes (SPCM-AQRH-14-FC-25690, Excelitas Technologies Corp., Canada). A multi-channel time-correlated single photon counting (TCSPC) device (OE-TCSPC-8 ps-6 channels, Simtrum Pte. Ltd., Singapore) allows us to store photon arrival times for each channel.

For the ATLAS-DCS system, which shares the same laser source, multiple-scattered light was collected by a multimode fiber with NA = 0.37 (MMF; HPCF1000/1100/1400-µm core/cladding/coating diameter, SC Intelligent Fiber Technology, Co. Ltd., Shenzhen, China) placed at $\rho$ from the source, and then coupled into the ATLAS camera. The fiber tip on the detection path was mounted on a fiber adapter (SM05SMA, Thorlabs) and attached to a 5-axis optic mount (K5X1, Thorlabs) to control the spacing between the tip and the camera sensor. The sensor was fixed on the optical table, and the illumination light intensity was adjusted to 30 mW, as verified by a power meter (TS2-TP100, Changchun New Industries Optoelectronics Technology, Co., Ltd., China). The laser spot diameter at the tissue surface is approximately 2 mm.

Fig. 2 shows a schematic of the conventional and ATLAS-DCS instrumentation. In the conventional DCS system, the scattered light is routed to the Excelitas sensor via a single-mode fiber, and a correlator counts the digital TTL pulses generated by the sensor to compute $g_2(\tau)$. In the ATLAS-DCS system, a printed circuit board (PCB) hosts an Opal Kelly XEM7310-A200 FPGA. Figs 2(b) and 2(c) show the photon count versus the number of pixels during no measurement and during tissue measurements in a dark room, respectively. Fig. 2(d) displays the probability density of photon counts; the average photon count is $208.52 \pm 31.29$.

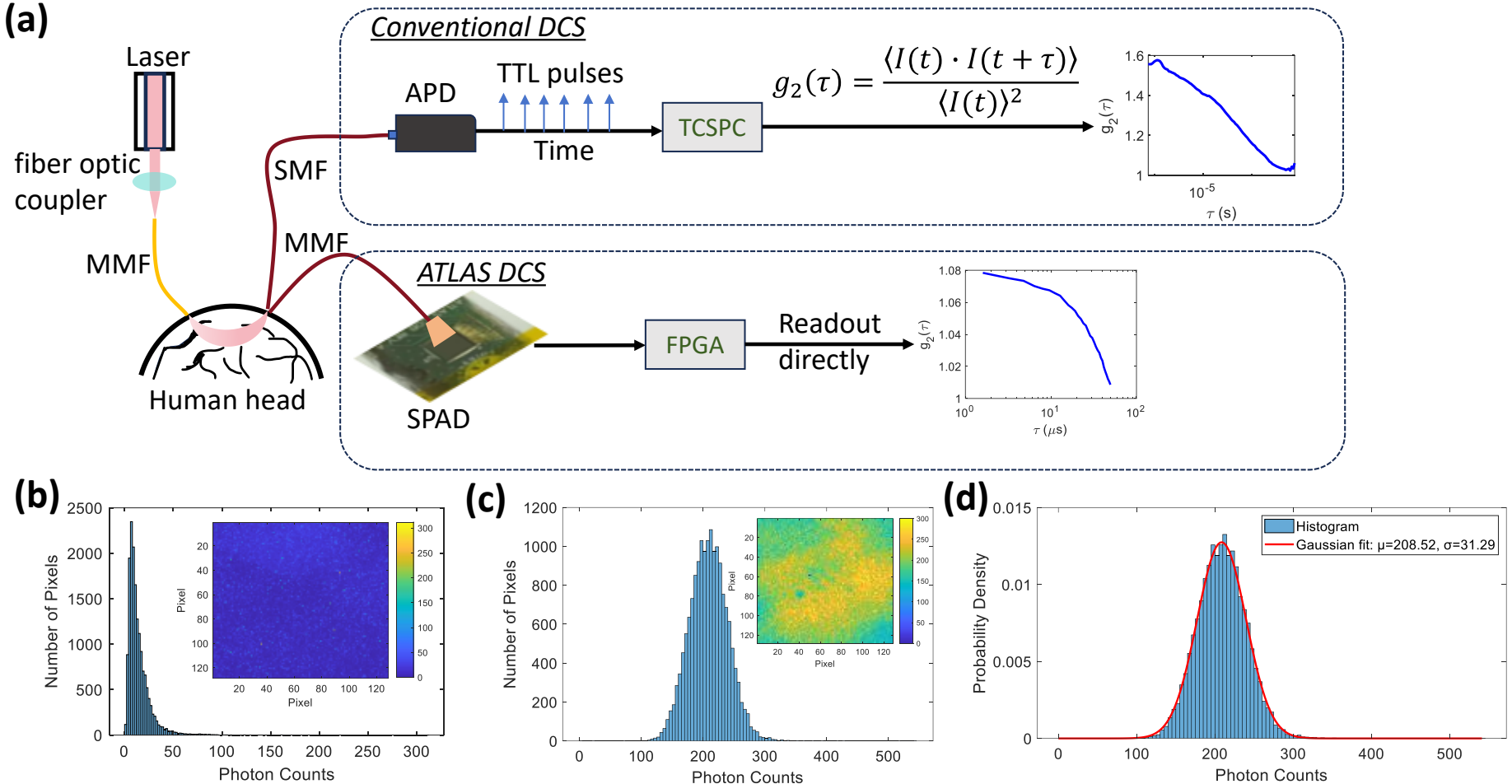

Fig. 2. (a) Schematic of the Conventional and ATLAS-DCS instrumentation in the tissues. A highly coherent, long-coherence-length laser illuminates the sample via a multimode fiber (MMF). The motion of red blood cells induces fluctuations in the intensity of backscattered light, collected at a separation distance ρ from the source. In the conventional DCS system, the scattered light is routed to the Excelitas sensor via a single-mode fiber (SMF), and a correlator counts the arrival of digital TTL pulses generated by the sensor to compute $g_2$. In the ATLAS-DCS system, a PCB board hosts an Opal Kelly XEM7310-A200 FPGA. (b) and (c) presents the photon counts Vs the number of pixels during no measurement and tissue measurements in a dark room, respectively. (d) is probability density of photon counts and average photon counts is 208.52±31.29.

### 2.4. *FD-NIRS Measurements*

To improve the estimation of DCS measured BFi, we independently measured the absorption and reduced scattering coefficients ($\mu_a$ and $\mu_s'$, respectively) of the tissue or phantom. And we used a customized commercial frequency-domain NIRS (FDNIRS) system (MetaOx, ISS Inc. USA) to estimate $\mu_a$ and $\mu_s'$ at eight wavelengths (673, 688, 702, 729, 748, 780, 801, and 830 nm). The system operated at a modulation frequency of 110 MHz and used four photomultiplier tube (PMT) detectors with gain modulation at 110 MHz + 5 kHz to achieve heterodyne detection at 5 kHz. The patient interface consisted of a custom 3D-printed FD-NIRS sensor with one source optode and four detector optodes placed at distances of 1.5, 2, 2.5, and 3.5 cm from the source, all housed in a black rigid holder. All NIRS source and detector optodes were custom 2.5-mm fiber bundles composed of 50-µm multimode fibers. For FD-NIRS data acquisition, the optical sensor was manually held over the left forearm and then over the forehead. AC amplitude and phase data were acquired for 60 seconds at a rate of 25 Hz. Each measurement was repeated three times. The fitted $\mu_a$ and $\mu_s'$ values obtained from the FD-NIRS module were used for subsequent DCS modeling. Since the 785 nm wavelength used in our DCS system was not included in the MetaOx dataset, we interpolated the $\mu_a$ and $\mu_s^{'}$ values at 785 nm.

## 2.5. *DCS Data Processing*

For conventional DCS, the photon count was recorded by the TCSPC module. They were then manually binned within a specified time window (e.g., $0.75\times 10^{-6}$ s) to obtain intensity time traces. These traces were used to compute the $g_2(\tau)$, where $\tau$ ranges from $7.5 \times 10^{-7}$ to $7.5 \times 10^{-4}$ s. Each $g_2(\tau)$ curve was then fitted to the semi-infinite solution of the correlation diffusion equation to extract BFi. For all fits, we assumed a refractive index of 1.37 and used the $\mu_a$ and $\mu_s^{'}$ coefficients at 785 nm measured from the corresponding object (phantom or tissue).

In contrast, the ATLAS-DCS system (Fig. 2(a)) integrates data acquisition and processing on-chip, enabling efficient ensemble-mode (or DCS imaging mode) computation of $g_2(\tau)$ at a high speed of 116 Hz, meaning that $g_2(\tau)$ is obtained directly. In our implementation, with a clock frequency of 20 MHz, $\tau$ ranges from 1.6 μs to 49.6 μs. The subsequent fitting procedure was the same as that used for conventional DCS, and the required optical parameters were identical.

## 2.6. *Phantom Experiments*

Fig. 3 shows the experimental setup for the phantom experiments. Before any DCS measurement, we measured the phantom as shown in Fig. 3(a) to obtain its optical properties. We 3D-printed a 10 cm × 10 cm × 10 cm open-topped black box to hold the phantom, which consisted of a mixture of water and 20% intralipid at a ratio of 1:30. The FD-NIRS probe was immersed at the phantom surface. After that, we performed measurements with both DCS systems on this homogeneous liquid phantom at room temperature (~ 25 °C), as shown in Fig. 3(b). Finally, we conducted temperature-variation experiments on the same phantom while gradually increasing its temperature from 15 °C to 26 °C. This temperature ramp increased the Brownian motion of scatterers in the phantom, allowing validation of the DCS-derived Db. As shown in Fig. 3(c), the phantom was placed in a glass beaker, which was then placed in a water-bath temperature controller. The temperature could be set manually on the controller's panel. We limited the temperature change to 1 °C per step and waited several minutes at each temperature to ensure even heating throughout the phantom. We used a custom fiber holder for both the single-mode fiber (conventional DCS) and the multimode fiber (ATLAS-DCS), with $\rho$ = 25 mm for both systems.

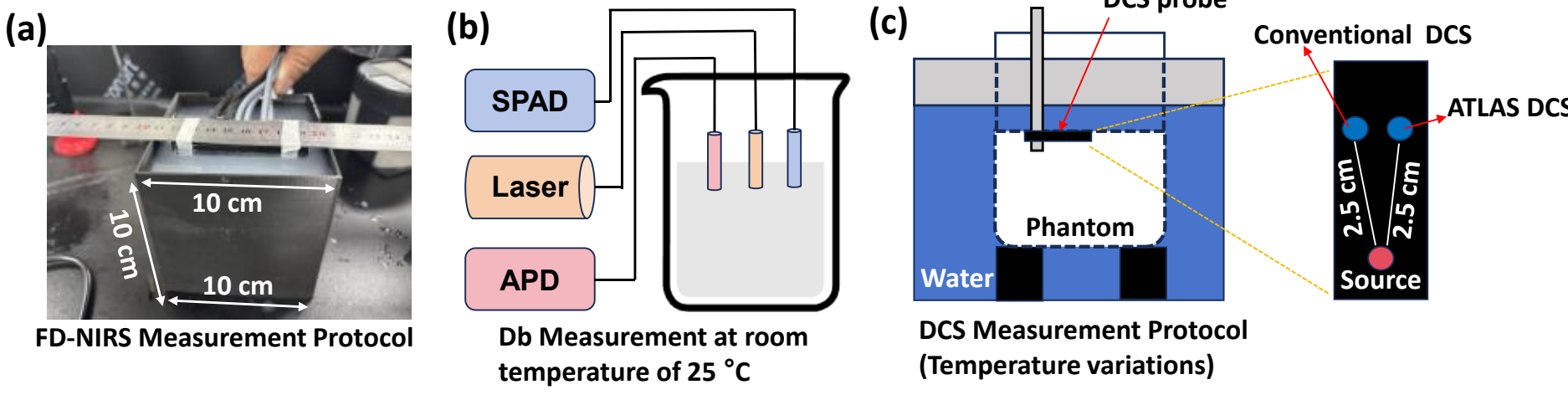

Fig. 3. FD-NIRS and DCS measurement protocols on phantom: (a) FD-NIRS protocol; (b) room-temperature Db measurement using ATLAS DCS and conventional DCS; (c) Two systems measurement with temperature variation on phantom.

## 2.7. *The Mental Arithmetic Protocol and Data Analysis*

Before each mental arithmetic (MA) session, subjects were instructed to sit still and keep their minds blank. A custom-built DCS optical probe was secured above the prefrontal cortex on the forehead. Two computers were used: one ran our in-house DCS acquisition software, and the other ran a MATLAB script with the open-source Psychtoolbox [46] for stimulus presentation. The computers communicated via a network socket. During the experiment, the stimulus computer sent real-time markers to the DCS acquisition computer at four points in each trial: rest onset (fixation cross appeared), MA onset (cross disappeared and a two-digit arithmetic problem appeared), MA offset (problem disappeared), and rest offset (cross reappeared). Each marker was timestamped and embedded directly in the DCS data stream. Each trial consisted of three consecutive 1-min blocks: rest (fixation cross only), MA (randomly generated two-digit subtraction or addition problems presented every 2 s; subjects performed mental calculation without verbal response), and rest again. The fixation cross was present during both rest periods to maintain a constant visual baseline. A total of two trials were completed on a single subject, with a 30-s break between trials.

After data acquisition, BFi was derived from the raw DCS intensity autocorrelation functions as described in Section 2.5. To remove periodic cardiac artifacts, we applied a cardiac signal removal algorithm to the entire continuous BFi time series. Specifically, systolic peaks in the BFi signal were automatically identified, and a mean cardiac waveform was constructed by ensemble averaging over all cardiac cycles. This mean cardiac signal was then subtracted from the original BFi, yielding a cardiac-cleaned BFi signal. The cleaned BFi was then segmented according to the recorded markers into individual trials, each containing the 1-min rest, 1-min MA, and 1-min rest periods. The resulting BFi responses were used for all further analyses.

## 2.8. *In vivo Measurements*

To compare the two DCS systems for measuring BF changes, we performed a forearm arterial occlusion experiment on the left arm of two healthy volunteers (one male, one female, of similar age). The aim of this measurement was to evaluate the performance of ATLAS-DCS. The probe, holding the emitter and detection fibers, was secured over the subject's flexor pollicis longus muscle, while a blood pressure cuff was placed on the upper arm. The measurement protocol consisted of a 20-s baseline, a 20-s cuff occlusion (230 mmHg), and a 20-s recovery after cuff deflation. Measurements were carried out on each subject on the same day; the probe was removed from the tissue surface when switching between subjects and then repositioned at the same location prior to measurements. To ensure safe illumination power for human tissue, the laser output was attenuated to deliver an average power of 30 mW to the skin surface. For each experiment, we computed the relative BFi ($rBFi = BFi/BFi_{baseline}$) by normalizing the measured BFi values to the mean baseline value. The protocol was approved by the Ethical

Committee of the University of Macau and conducted in accordance with the Declaration of Helsinki. All subjects gave written informed consent prior to measurement.

## 3. Results

### 3.1. *System validation using liquid phantoms*

To validate our systems, a liquid phantom was prepared by mixing 20% intralipid suspension with water at a ratio of 1:30, yielding a scattering coefficient of 6.05 $cm^{-1}$ at 830 nm. The phantom had no additional absorption beyond that of water.

Fig. 4(a) presents the $g_2(\tau)$ curves obtained simultaneously using the protocol in Fig. 3(b). The low $\beta$ (~0.1) is attributed to the use of multimode fiber combined with a SPAD detector, which reduces the coherence factor because the multimode fiber supports many independent spatial modes whose superposition reduces the contrast of the intensity autocorrelation, and because each macropixel averages over multiple uncorrelated speckle grains. We fitted the $g_2(\tau)$ curves to the semi-infinite DCS model using $\mu_a = 0.024$ $cm^{-1}$ and $\mu_s' = 6.66$ $cm^{-1}$ (interpolated to 785 nm). Other fitting parameters included $\rho = 25$ mm and a refractive index of 1.37. For the conventional DCS system, we used an integration time of 0.1 s and a time window of 0.75 μs.

We next compared the performance of ATLAS-DCS with that of conventional DCS and a commercial DCS system. SNR was calculated as mean(BFi)/std(BFi). The $\alpha \cdot D_b$ values for ATLAS DCS, conventional DCS, and commercial DCS (MetaOx ISS) were $1.30 \times 10^{-6} \pm 0.0073 \times 10^{-6}$ $mm^2/s$, $1.34 \times 10^{-6} \pm 0.091 \times 10^{-6}$ $mm^2/s$, and $1.6 \times 10^{-6} \pm 0.34 \times 10^{-6}$ $mm^2/s$, respectively. The corresponding SNRs were 178.38, 14.70, and 4.64, representing a 12-fold and 38-fold SNR improvement of ATLAS-DCS over conventional DCS and commercial DCS, respectively. Fig. 4(c) shows a Bland-Altman plot for $BFi_{ATLAS\text{-}DCS}$ versus $BFi_{conventional\ DCS}$. A negative trend was observed, indicating that the difference between the two measurements increased with the average value. The results of the temperature-ramped homogeneous phantom experiment are shown in Fig. 4(d) and 4(e).

For both systems, BFI increased monotonically with rising temperature. Specifically, conventional DCS yielded BFi values from $1.31 \times 10^{-6}$ $mm^2/s$ at 14 °C to $1.63 \times 10^{-6}$ $mm^2/s$ at 25 °C, while ATLAS DCS produced values from $1.37 \times 10^{-6}$ $mm^2/s$ to $1.51 \times 10^{-6}$ $mm^2/s$ over the same temperature range. These observations are consistent with the expected enhancement of scatterer Brownian motion at elevated temperatures and demonstrate the responsiveness of both DCS systems to temperature variations [47].

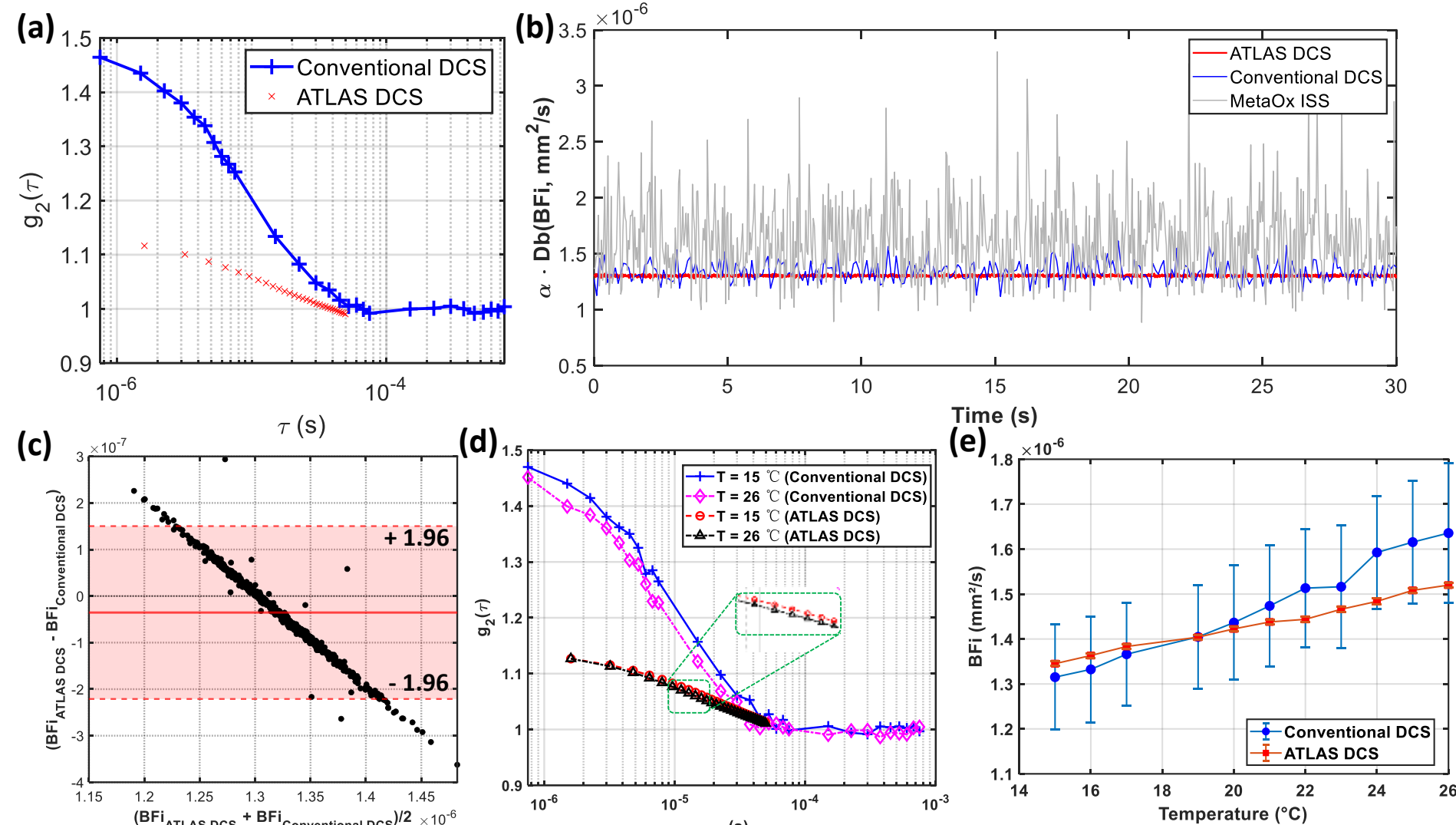

Fig. 4. (a) $g_2(\tau)$ curves for conventional DCS and ATLAS-DCS. (b) $\alpha D_b$ measured by conventional DCS, ATLAS-DCS, and MetaOx. (c) The Bland-Altman plot for $\text{BFi}_{\text{ATLAS-DCS}}$ and $\text{BFi}_{\text{conventional DCS}}$. The gray shaded region represents the 95% confidence interval for agreement. A negative trend is evident, indicating that as the average value increases, the difference between $\text{BFi}_{\text{ATLAS-DCS}}$ and $\text{BFi}_{\text{conventional DCS}}$ increases. (d) Representative $g_2(\tau)$ curves at different temperatures (15℃ and 26℃) from ATLAS DCS and conventional DCS (e) Comparison of BFindex (BFi, $\times 10^{-6}$ mm²/s) between Conventional DCS and ATLAS DCS as a function of temperature (14–26 °C). BFi increases with increasing temperature for both methods.

### 3.2. *Forearm Occlusion*

Fig. 5 shows the results of cuff occlusion measurements on two healthy subjects; the corresponding optical parameters are listed in Table 1 in Appendix. The relative blood flow index (rBFi) changes are shown for ATLAS-DCS (blue) and conventional DCS (red). Absolute BFi values were estimated over the entire measurement duration. To obtain rBFi, the absolute values were normalized to the baseline mean (first 20 s of the measurement). As shown in Fig. 5, a consistent trend was observed across all experiments: during cuff occlusion, rBFi rapidly decreased to a minimum, followed by a pronounced hyperemic peak after cuff release. This characteristic response was captured by both DCS systems at $\rho = 25$ mm and $\rho = 35$ mm for both subjects. Although there was a discrepancy between the two subjects, this difference may be attributed to sex-related physiological factors, such as baseline vascular tone, muscle mass, or sympathetic nervous system activity. The post-occlusion recovery kinetics were similar between the two systems, demonstrating good agreement in tracking dynamic flow changes despite differences in absolute amplitude.

In addition to the occlusion-induced dynamics, a notable difference in signal quality was observed at the larger source-detector separation ($\rho = 35$ mm). As shown in Figs. 5(b) and (d), the ATLAS-DCS system clearly resolved cardiac pulsations (beat-to-beat oscillations) in the rBFi trace during baseline and recovery phases. In contrast, the conventional DCS system at the same separation exhibited no discernible pulsatile pattern, with the signal appearing smooth and lacking cardiac rhythm. This advantage of ATLAS-DCS was consistent across both subjects, indicating that its improved signal-to-noise ratio enables the detection of fast hemodynamic oscillations at larger $\rho$ where conventional DCS fails, as shown in the zoomed-in Figs. 6(c) and (d).

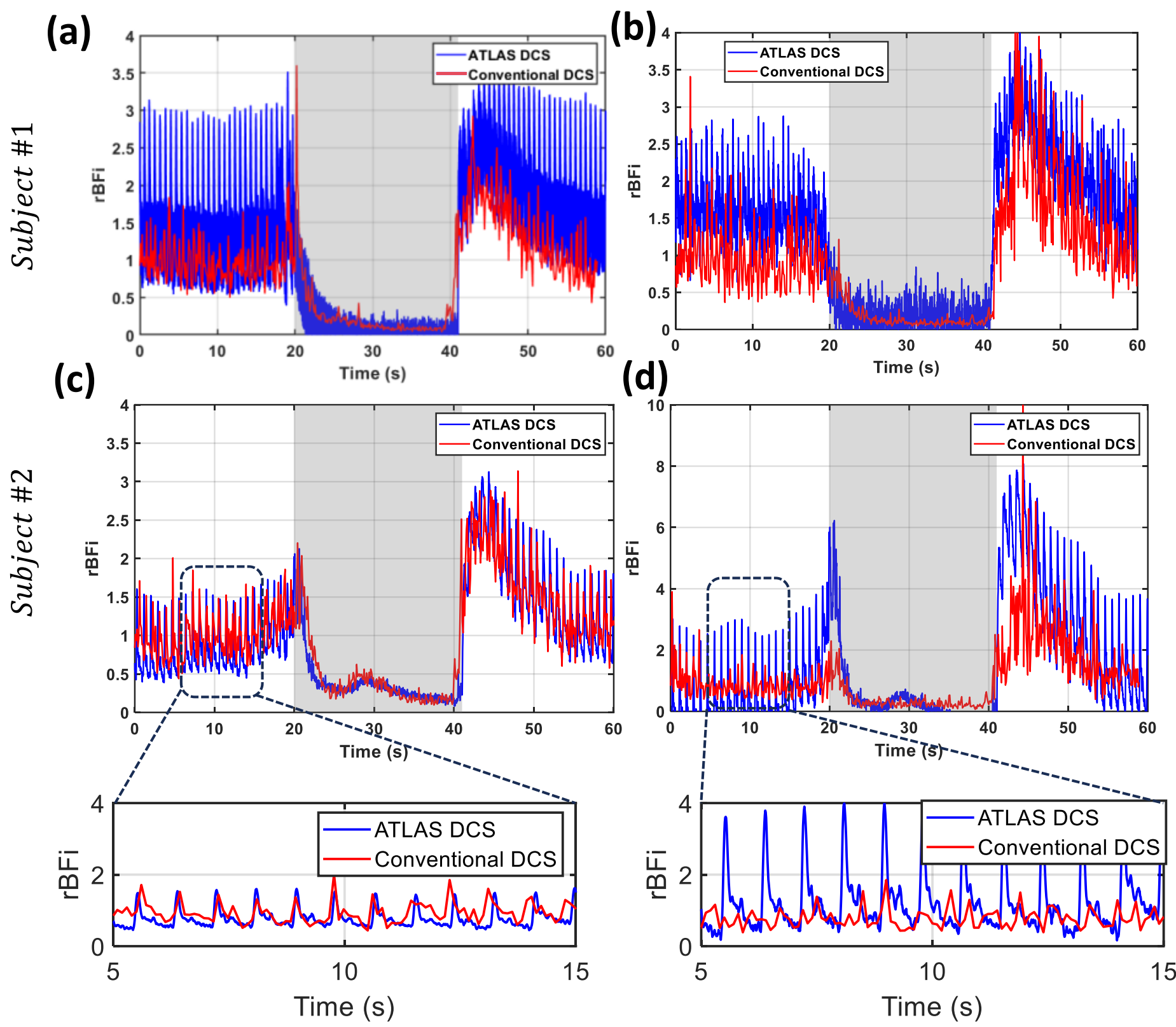

Fig. 5 The rBFi obtained by two DCS systems during cuff occlusion measurements on the forearms of two healthy adults. (a) and (b) are for Subject #1 at $\rho$ = 25 mm and $\rho$ = 35 mm, respectively. (c) and (d) are for Subject #2 at $\rho$ = 25 mm and $\rho$ = 35 mm, respectively, with their corresponding subsets shown in the range of 5 s to 15 s.

In order to demonstrate that the ATLAS-DCS and FD-NIRS devices can operate simultaneously, we further performed a forearm arterial occlusion protocol. The DCS and FD-NIRS probes were placed as close as possible to each other on the subject's left arm to ensure overlapping tissue sampling, and a blood pressure cuff was wrapped around the upper arm. After recording a 20-second baseline, the cuff was rapidly inflated to a pressure exceeding systolic blood pressure (~230 mmHg). The occlusion was sustained for 20 seconds and then abruptly released. The resulting time courses of total hemoglobin, deoxyhemoglobin (HbR), oxyhemoglobin (HbO), oxygen saturation ($SO_2$), and the DCS-derived rBFi for two subjects are displayed in Fig. 6.

Throughout the 20-second occlusion period, rBFi fell to near-zero levels, confirming effective blockade of arterial inflow. At the same time, HbO concentration showed a progressive decline, whereas HbR concentration increased, resulting in a drop in $SO_2$. After cuff deflation, a clear reactive hyperemia was observed: rBFi rose above baseline, HbO exhibited an overshoot, and HbR decreased below its baseline value. Regarding $SO_2$, it decreased during occlusion and, following cuff release, rose above the pre-occlusion level due to hyperemic washout, then gradually returned toward baseline. These response patterns were highly consistent across subjects, although some variation was noted in baseline values and the amplitude of changes. This finding is consistent with previous reports (Ref. [27]). In all participants, BF was effectively halted, and the reduction in $SO_2$ was approximately 3%.

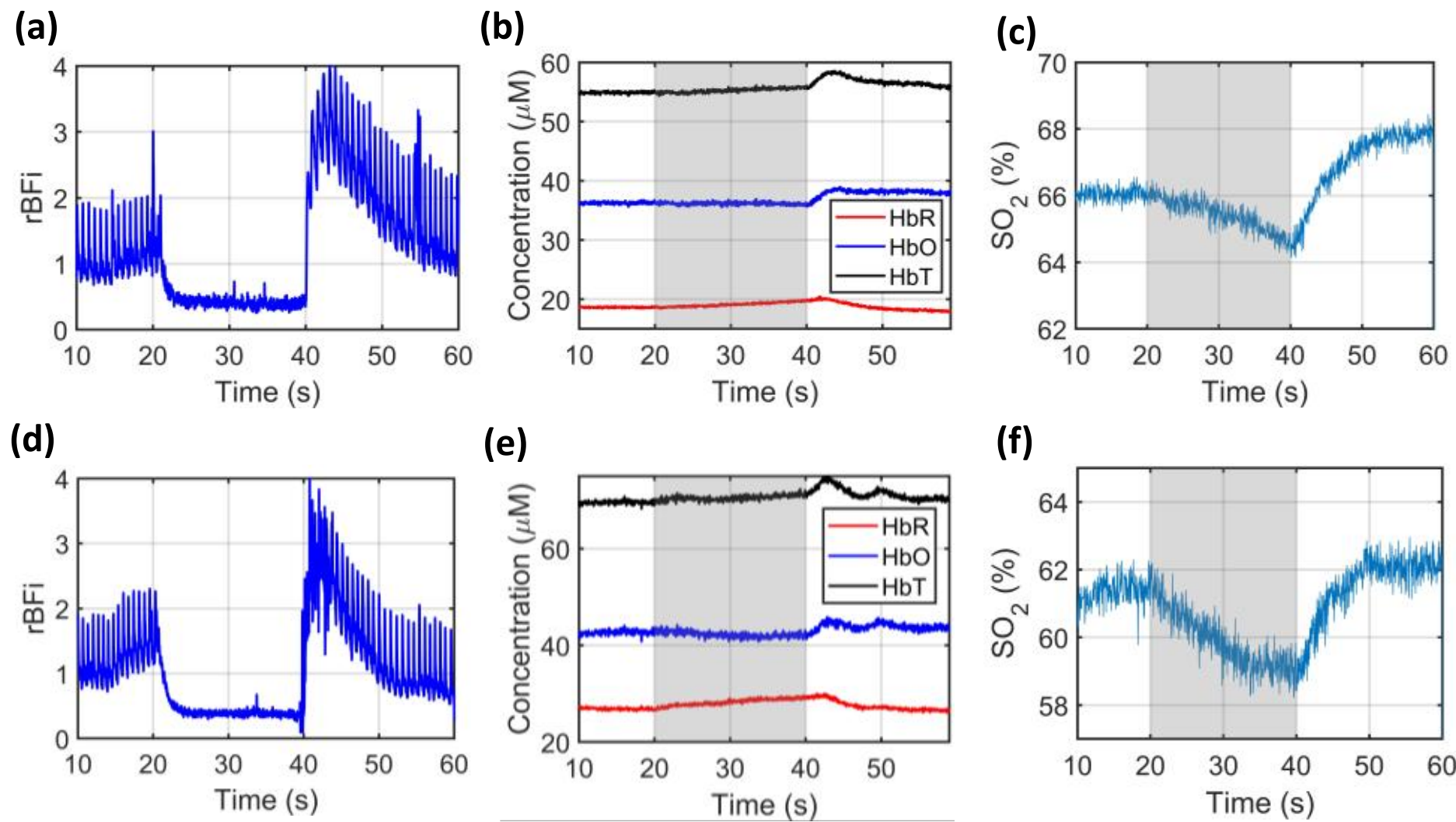

Fig. 6 Example forearm occlusion data from two subjects using combined ATLAS-DCS and FD-NIRS. (a, d) DCS-derived rBFi; (b, e) concentrations of HbO, HbR, and HbT; (c, f) $SO_2$ traces, shown before, during, and after the 20-second occlusion. The blood pressure cuff was inflated to 230 mmHg.

### 3.3. *Forehead*

Fig. 7 presents the results of forehead measurements on two healthy subjects; the corresponding optical parameters are listed in Table 2. Clear pulsatile BFi traces were obtained at $\rho = 45$ mm by fitting $g_2(\tau)$ to a semi-infinite homogeneous model. The quality of the pulsatile waveform appeared comparable to that of state-of-the-art functional interferometric diffusing wave spectroscopy (fiDWS)/speckle contrast optical spectroscopy (SCOS) reported by Kim *et al*. at $\rho = 35$ mm [48,49], which used a relatively complicated optical setup or pulse strategy. Even at $\rho = 45$ mm, the pulsatile waveform quality was sufficient to enable self-alignment of BFi traces with respect to the cardiac cycle. At $\rho = 50$ mm, the heart rate (HR) was still resolvable in the fast Fourier transform (FFT) spectrum (Fig. 7(d)). Fig. 7(b) shows the average BFi over 10 seconds as a function of $\rho$. BFi increased monotonically with $\rho$, attributed to the progressively larger contribution from deeper cerebral tissue (where BFvelocity is higher) and reduced sensitivity to superficial scalp hemodynamics. Although this trend was clearly observed, the error bars were largest at $\rho = 50$ mm, likely due to the lower SNR at this separation. Moreover, at such a large $\rho$, laser decoherence during the measurement window must be accounted for in quantitative analysis. Therefore, a $\rho$ value over 50 mm was not pursued further.

Here, we defined SNR as $10 \log_{10}(P_{signal} / P_{noise})$ dB, where $P_{signal}$ is the total power in the cardiac frequency band (0.8–2.5 Hz) and $P_{noise}$ is the total power in the remaining frequency bands (0–0.8 Hz and 2.5–10 Hz), obtained from the power spectral density of the detrended BFi time series. SNR as a function of ρ exhibited a unimodal trend, peaking at $\rho = 35$ mm for both measurement configurations.

Specifically, the SNR (solid line) increased from 2.8 dB at $\rho = 20$ mm to 4.9 dB at 35 mm, then decreased to 3.9 dB at 40 mm and 1.9 dB at 50 mm. The fitted SNR (dashed line) followed a similar pattern.

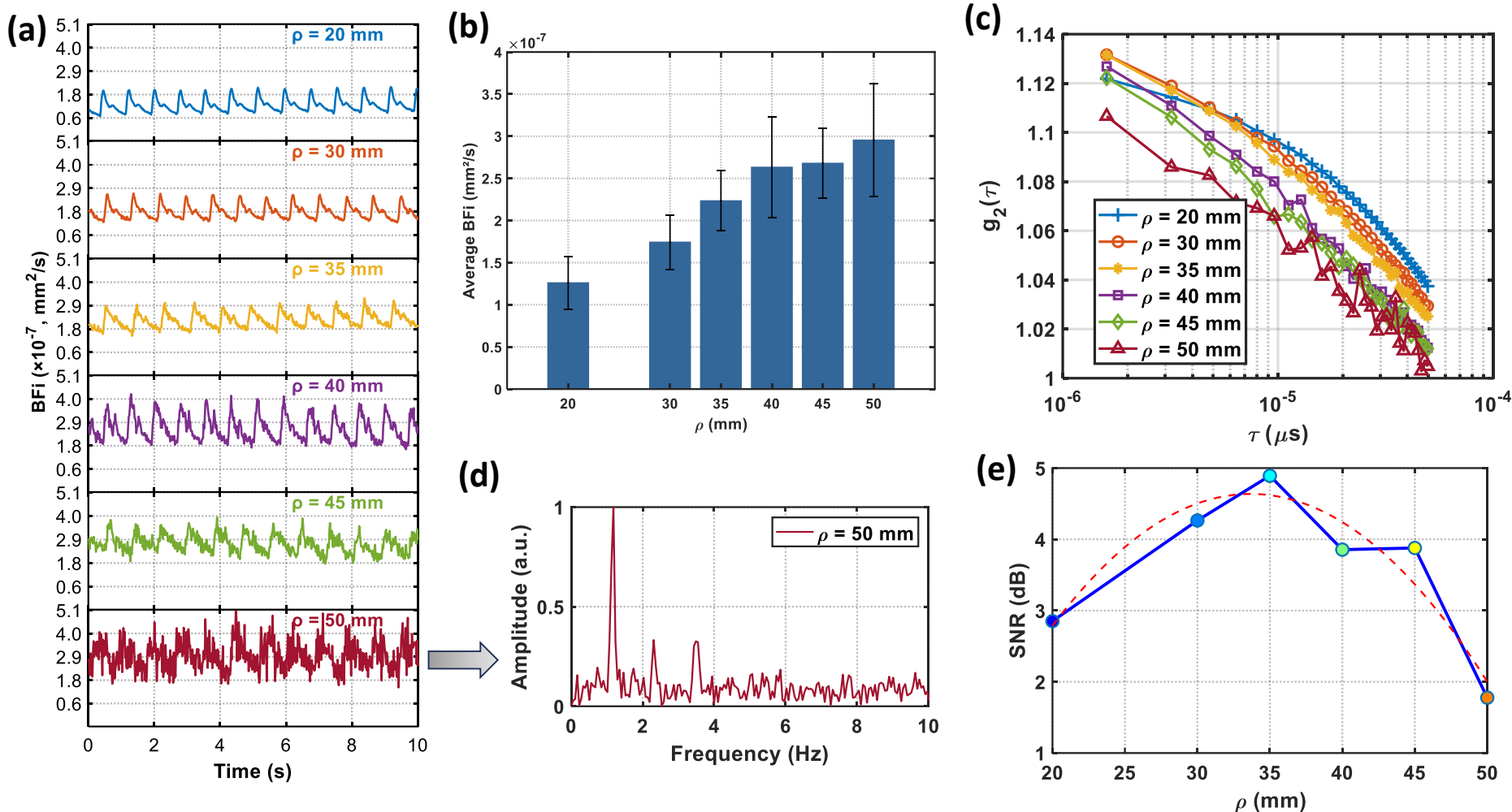

Fig. 7 ATLAS DCS monitors pulsatile BFi at $\rho$ = 20–50 mm. (a) Pulsatile BFi traces from a single subject with $\rho$ = 20-50 mm. (b) the averaged BFi within 10 seconds versus $\rho$, and the error bars indicate the standard deviation. (c) $g_2(\tau)$ at $\rho$ = 20–50 mm for the same Subject in (a). (d) The fast Fourier transform (FFT) spectrum of the pulsatile BFi trace at $\rho$ = 50 mm. (e) SNR vs. $\rho$.

### 3.4. *ATLAS DCS for brain activation*

Finally, we performed human brain function measurements during a mental arithmetic task at $\rho$ = 30 for a single participant across two repeated trials, as shown in Fig. 8. As described in the Methods section, we used the open-source Psychtoolbox to control stimulus timing. Math problems were presented randomly to prevent anticipation, as illustrated in Fig. 8(a). Figs. 8(b) and (e) show the raw signal from the ATLAS DCS system, and Figs. 8(c) and (f) show the cardiac signal. The zoomed-in subplot (Figs. 8(h) and 8(i)) clearly reveals cardiac pulsatility. After removing the cardiac signal, we observed CBF changes induced by brain activation, as shown in Figs. 8(d) and 8(g). Across different trials for a given subject, BF increased significantly (by 7–16%) during activation and returned to baseline afterwards, consistent with literature values [48]. The shaded area in Fig. 8(d) and 8(g) highlights the activation period.

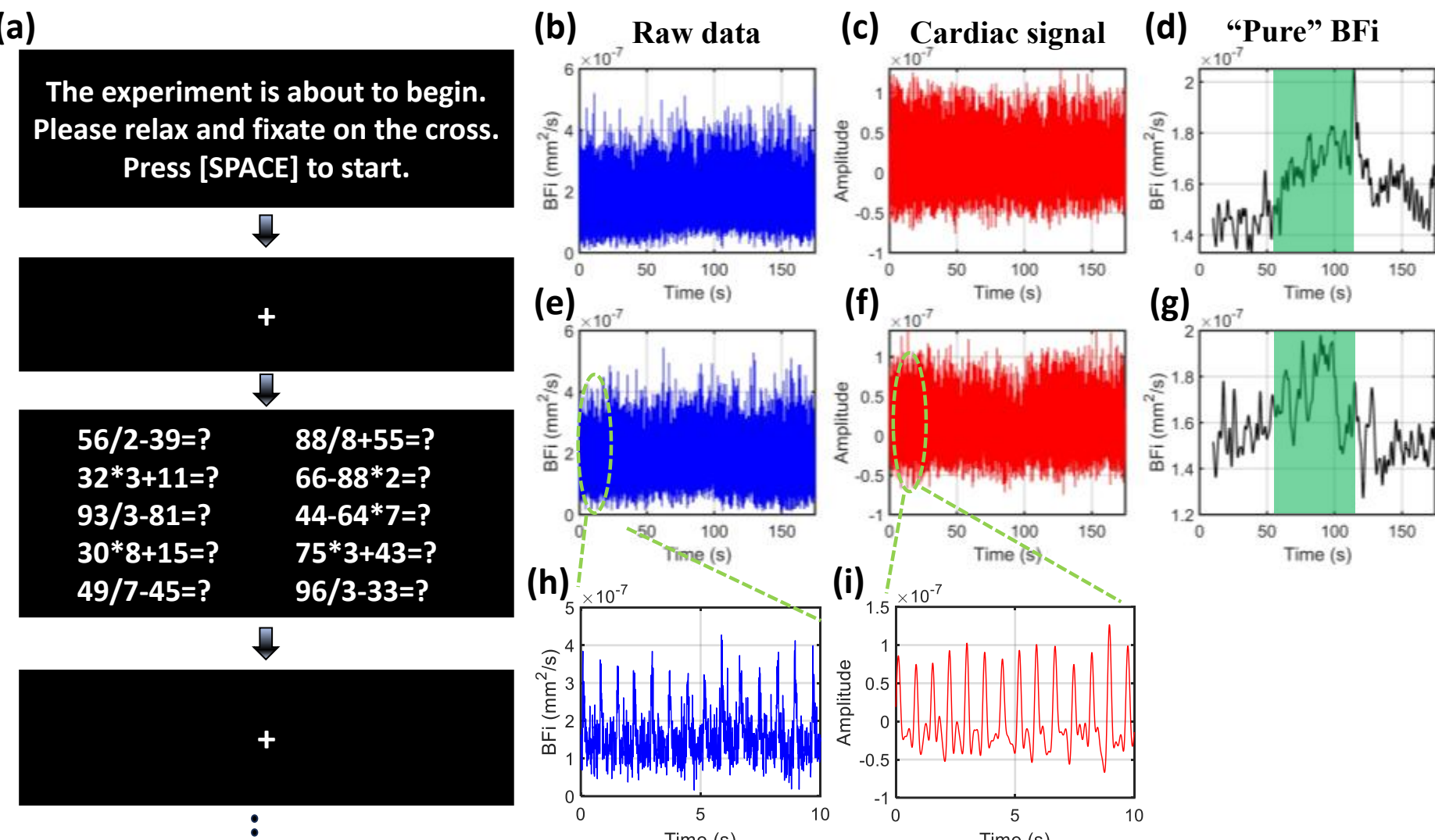

Fig. 8. ATLAS-DCS during prefrontal cortex activation. (a) Protocol of the mental arithmetic (MA) experiment. During rest, a fixation cross ("+") is displayed; during the task, arithmetic questions (e.g., "56/2-39=?") are presented for calculation. (b), (e), and (h) Raw data recorded during the experiment. (c), (f), and (i) cardiac signal (amplitude vs. time). (d) and (g) BFi signal ($mm^2/s$ vs. time). Both signals show changes associated with prefrontal cortex activation.

## 4. Discussions

A persistent trade-off in optical BF monitoring lies between achieving sufficient brain-to-scalp sensitivity (requiring large ρ) and maintaining adequate SNR at those ρs. Several recent advances have attempted to break this trade-off. Carp *et al.* [49] demonstrated DCS at 1064 nm, where lower tissue absorption and higher maximum permissible exposure increase photon budget, while the slower autocorrelation decay shifts the signal to lower-noise delay bins. However, 1064 nm DCS still relies on expensive superconducting nanowire single-photon detectors and does not fundamentally increase the number of parallel speckle channels. Zhou *et al.* [50] and Xu et al. [51] developed interferometric DWS (iDWS), which uses heterodyne gain to achieve shot-noise-limited detection with a CMOS line-scan camera, enabling pulsatile BFi measurements at $\rho = 35$ mm and beyond. Despite its high sensitivity, iDWS requires a complex Mach–Zehnder interferometer and a multimode fiber sample arm that is sensitive to motion and vibrations, limiting its robustness for wearable or long-term monitoring. Separately, multi-speckle DCS and SCOS have leveraged SPAD arrays or CMOS cameras to parallelize detection over hundreds to thousands of speckles [48,52]. These methods drastically improve SNR at moderate ρ (~30 mm), but most implementations rely on off-chip processing (e.g., saving raw photon streams or images) and have not yet demonstrated real-time, high-frame-rate BFi extraction at $\rho > 40$ mm with simple fiber-based optics.

In contrast, the ATLAS-DCS system presented here combines three key advantages: (i) on-chip, hardware-accelerated autocorrelation using a 31-tap shift-register architecture embedded in each macropixel, which eliminates the need for off-chip correlation post-processing; (ii) massively parallel detection across a 128 × 128 macropixel array, each integrating 16 SPADs, providing a total of up to 16,384 effective detection channels; and (iii) operation at 785 nm using a low-cost multimode fiber collection path, without interferometry or pulsed illumination. As a result, ATLAS-DCS achieves real-time BFi acquisition at 116 Hz, resolves cardiac pulsations at $\rho = 45$ mm, and detects brain activation during a mental arithmetic task at $\rho = 30$ mm, all with a simple, stable, and potentially wearable optical probe. Compared with 1064 nm DCS, our system avoids expensive detectors and does not suffer from slower decorrelation that challenges fitting at short lag times. Compared with iDWS, ATLAS-DCS is optically simpler, less sensitive to fiber motion, and directly outputs autocorrelation curves without heterodyne stabilization. Compared with multi-speckle DCS/SCOS, our on-chip correlation enables true real-time processing (no frame storage or post-hoc correlation), while maintaining high SNR at large ρ through the combination of high pixel count and efficient on-chip averaging. Thus, ATLAS-DCS offers a practical, high-throughput, and low-cost pathway to wearable, real-time cerebral BF monitoring.

The development of SPAD array cameras in recent years has been a key driver of progress in DCS. Array formats have expanded from 5 × 5 to 512 × 512 pixels, and such detectors have been shown to deliver substantial gains in SNR. However, most existing SPAD-based DCS systems are restricted to offline data analysis and lack real-time detection capability. The ATLAS sensor addresses this limitation as, to the best of our knowledge, the first chip-based autocorrelation solution. By eliminating the sensor-to-FPGA I/O bottleneck, it supports parallel computation across the full array down to the shortest lag times, enabling the highest achievable sampling rates. As illustrated in Fig. 7(a), our ATLAS-DCS system continuously measures pulsatile BFi from the human forehead at $\rho$ = 20-50 mm with a sampling rate of 116 Hz. For optical BFi monitoring, these results offer an unprecedented combination of temporal resolution and brain-to-scalp sensitivity. In a series of MA experiments, we captured the time course of functional hyperaemia during the task, including its pulsatile component. To the best of our knowledge, these represent the highest-speed BF measurements reported at such temporal resolution, providing new insight into flow physiology and opening up opportunities for cerebral health monitoring. Our pulsatile BFi waveforms clearly resolve key features of the entire cardiac cycle, including the dicrotic notch, demonstrating data quality comparable to high-speed arterial line recordings. The high temporal resolution achieved here holds clear promise for the assessment of cerebrovascular compliance. Comparison of flow pulsatility measured at the arm and the brain, for example, could reveal differences in arterial elasticity, while measurements of cerebrovascular compliance may help identify arterial stiffening or dissection, features implicated in conditions including atherosclerosis, cerebral amyloid angiopathy, and stroke.

As shown in Fig. 6, our ATLAS-DCS system combines FD-NIRS and DCS into a single measurement, which we believe represents an effective integration of these two techniques [27,53]. By enabling simultaneous oximetry and flow measurements, it allows quantification of the tissue oxygen metabolic rate ($MRO_2$) a parameter closely linked to underlying physiological and pathological states [54]. This measure is particularly important in the brain (cerebral $MRO_2$, i.e., $CMRO_2$), given the organ's heavy dependence on aerobic metabolism [39]. Numerous studies using combined NIRS-DCS for $CMRO_2$ assessment have been reported in clinical neuromonitoring, neonatal development, and functional brain research[17,18,22]. Therefore, further progress in this field using large SPAD arrays would greatly benefit from a unified device that performs parallel NIRS and DCS measurements with a single user interface.

Nevertheless, the proposed chip-based correlator has certain limitations. It is currently restricted to 31 autocorrelation points. The minimum lag time is set by the pixel clock; at our chosen 20 MHz, $\tau$ spans 1.6 μs to 49.6 μs, which is sufficient for the physiological signals reported here. Compared with to CMOS sensors used in fiDWS [50,55] and SCOS [48], SPADs remain relatively expensive, a factor that currently constrains their scalability. However, with ongoing progress in silicon manufacturing and the potential for mass production at scale, this cost disadvantage is expected to narrow over time and is unlikely to constitute a lasting barrier to widespread adoption. A further limitation relates to the forearm cuff-occlusion experiments: the optical probe was not co-located with the other modalities, so each may have sampled a different tissue region. Future studies will co-register measurement positions to enable more rigorous cross-modality comparison.

## 5. Conclusion

We have characterized the ATLAS-DCS system for assessing CBF up to $\rho = 50$ mm, benefited from on-chip embedded autocorrelators. It offers real-time BF monitoring at a 116 Hz sampling rate while maintaining the accuracy and performing better than a conventional DCS system. These preliminary results suggest the system has the potential to inform assessments of cerebrovascular reactivity, though further validation is required. We demonstrated that the device could assess cerebrovascular reactivity. We believe that it has potential applications in assessing cerebrovascular diseases and in brain research.

## Data and Code Availability

All datasets and codes supporting the findings of this study are available from the corresponding author upon reasonable request.

## Funding

This work has been funded by: the Engineering and Physical Sciences Research Council (Grant No. EP/T00097X/1 and No. EP/T020997/1), the Quantum Technology Hub in Quantum Imaging (QuantiC), the University of Strathclyde, and James Anderson Charitable Trust (ABBIE); the Economic Development Bureau of Hengqin (No. 2430004045063); the Macao Science and Technology Development Fund (FDCT 0014/2024/RIB1); the STI2030-Major Projects (No. 2022ZD0213300); and the University of Macau (MYRG-GRG2023-00038-FHS-UMDF, MYRG2022-00054-FHS, and MYRG-GRG2024-00259-FHS).

## Acknowledgments

The authors extend their gratitude to the Diffuse Correlation Spectroscopy (DCS) community for the valuable discussions and collaborations over the past two years, particularly with esteemed scholars from Linköping University, Harvard Medical School, Western University, Duke University, Boston University, and Singapore A*STAR. We are especially grateful to Johannes D. Johansson and Stefan A. Carpfor their insightful discussions that contributed to this work. Additionally, we acknowledge Saeed Samaei, Mamadou Diop, Lucas Kreiss, Byungchan Kim, Louis Gagnon, Wenhui Liu, Tom Y. Cheng, Mitchell B. Robinson and Xiaojun Chen for their valuable discussions in the field of diffuse optics over the past three years. We also thank Dennis M. Hueber for his help in FD-NIRS measurements. ATLAS was designed in a project funded by Reality Labs, Meta Platforms Inc., Menlo Park, CA 94025, USA. We are grateful to STMicroelectronics for CMOS manufacturing of the device within the University of Edinburgh Collaboration Agreement.

**Conflicts of Interest**

The authors declare that there are no conflicts of interest relevant to this article.

## Appendix 1: Comparison of single-exponential and semi-infinite diffuse correlation spectroscopy models for BFi estimation

In DCS experiments, the dynamic scattering media usually consist of scatterers exhibiting either random flow (ballistic) or Brownian (diffusive) motions. To obtain the decorrelation lifetime [56], $\tau_c$, a simpler functional expression is preferred for fitting measured $g_2(\tau)$:

$$f(\tau) = 1 + a \cdot exp(-\nu\tau). \qquad \text{(A1)}$$

We define $\tau_c = 2/\nu$, where the coefficient $\nu$ of the exponential term is referred to as the "decorrelation speed" as adopted from Liu *et al.* [31]. The decorrelation speed, directly related to BF velocity, reveals the pulsatile nature of the flow, indicated by periodic peaks in the time series. The BFi is proportional to $1/\tau_c$ [31]. Variations to τ_c can then be directly attributed to, for example, a change in blood flow.

Fig. A1 shows both the single exponential model and semi-infinitely DCS model were applied to DCS data acquired at $\rho$ = 20 mm and 30 mm. Fig. A1(a) shows the measured $g_2(\tau)$ for both separations. The curve at $\rho$ =30 mm decays more slowly than that at ρ =20 mm, indicating that photons collected at a larger $\rho$ have undergone more scattering events and therefore sample deeper tissue layers. This slower decorrelation is consistent with the presence of slower flow components in deeper microvasculature (e.g., cerebral parenchyma), whereas the faster decay at $\rho$ = 20 mm reflects the contribution of superficial, faster-moving blood (e.g., scalp and skull) [17]. The distinct shapes of the two $g_2(\tau)$ curves underscore the necessity of using multi-distance DCS when depth discrimination is required [57]. Figs. A1(b) and (c) compare the normalized BFi derived from the single-exponential model ($1/\tau_c$, red) and the semi-infinite model ($BFi_{semi}$, blue) at $\rho = 20$ mm and $\rho = 30$ mm, respectively. Overall, excellent agreement is observed between the two models: both time series exhibit nearly identical temporal fluctuations, including the positions of peaks and troughs, as well as the long-term trends. This strong correspondence indicates that both the empirical exponential fit and the physically motivated semi-infinite solution capture the same underlying microvascular flow dynamics with comparable fidelity [16,17].

Nevertheless, two subtle differences can be noted. First, at $\rho$ = 20 mm (Fig. A1(b)), the exponential model yields slightly larger amplitude oscillations than the semi-infinite model, which is attributed to the higher sensitivity of the exponential fitting to fast decorrelation components originating from superficial tissues [3]. Second, at $\rho$ = 30 mm (Fig. A1(c)), the two traces become nearly indistinguishable, suggesting that the semi-infinite model performs optimally at a larger ρ where the homogeneous medium approximation is more valid. Overall, the close match between the two models across both separations validates the robustness of our data processing and supports the use of either method for relative BFi monitoring in similar experimental settings. Indeed, the exponential model offers a straightforward fitting procedure with minimal computational cost and is particularly sensitive to rapid decorrelation events associated with pulsatile blood flow. However, because it reduces the complex tissue dynamics to a single decorrelation process, the exponential model does not account for contributions from distinct scattering populations (e.g., shallow vs. deep tissue layers). The semi-infinite model, by contrast, incorporates the physical propagation of photons in a scattering medium and has been widely adopted for quantitative BFi estimation in homogeneous tissue geometries.

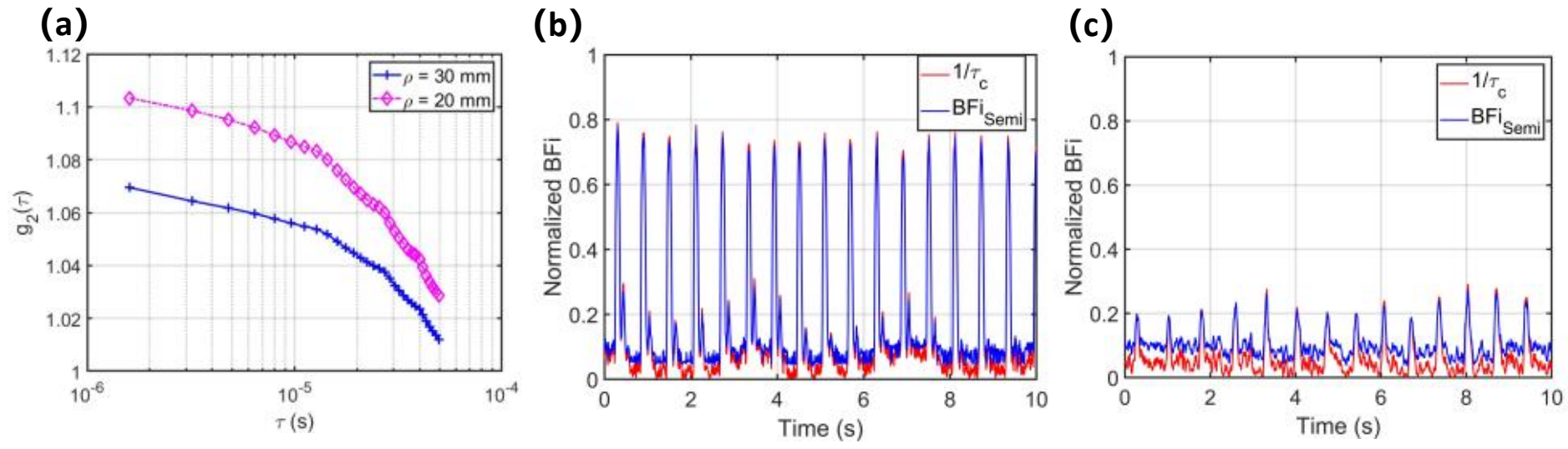

Fig. A1. (a) $g_2(\tau)$ curves at $\rho = 20$ mm and $\rho = 30$ mm. (b) BFi curves derived from the single-exponential model (red, $1/\tau_c$) and the semi-infinite model (blue, $BFi_{\text{Semi}}$) at $\rho = 20$ mm. (c) Same as (b) but at $\rho = 30$ mm.

## Appendix 2: Conventional DCS data Preprocessing pipeline

Fig. A2 shows the conventional DCS data analysis pipeline. When the light travel through sample (e.g., human tissues), the scattered light was detected by the photon counting APD, which generates a stream of digital TTL pulses, directly fed into a TCSPC module. Inside the TCSPC module, a high-precision time-to-digital converter (TDC) is integrated and runs continuously with an internal reference clock (25 MHz). When the rising edge of a photon pulse arrives, the TDC immediately captures the current value of its internal counter. This captured value is the macro time stamp of that photon, representing the time interval from the start of the measurement to the arrival of the photon. Each time stamp is encoded as a 64-bit binary integer (which includes channel information) and is transmitted by the TCSPC module to the computer via a USB interface, then saved in binary format. In subsequent data processing, these binary time stamps are parsed to obtain the statistical distribution of photon arrival times, which is then used for photon-counting analysis or for computing the second-order correlation function (e.g., $g_2(\tau)$), a key step in DCS measurements. As illustrated in Fig. A2, the raw macro time stamps are binned into a histogram (e.g., time window = $0.75 \times 10^{-6}$ s) for further processing.

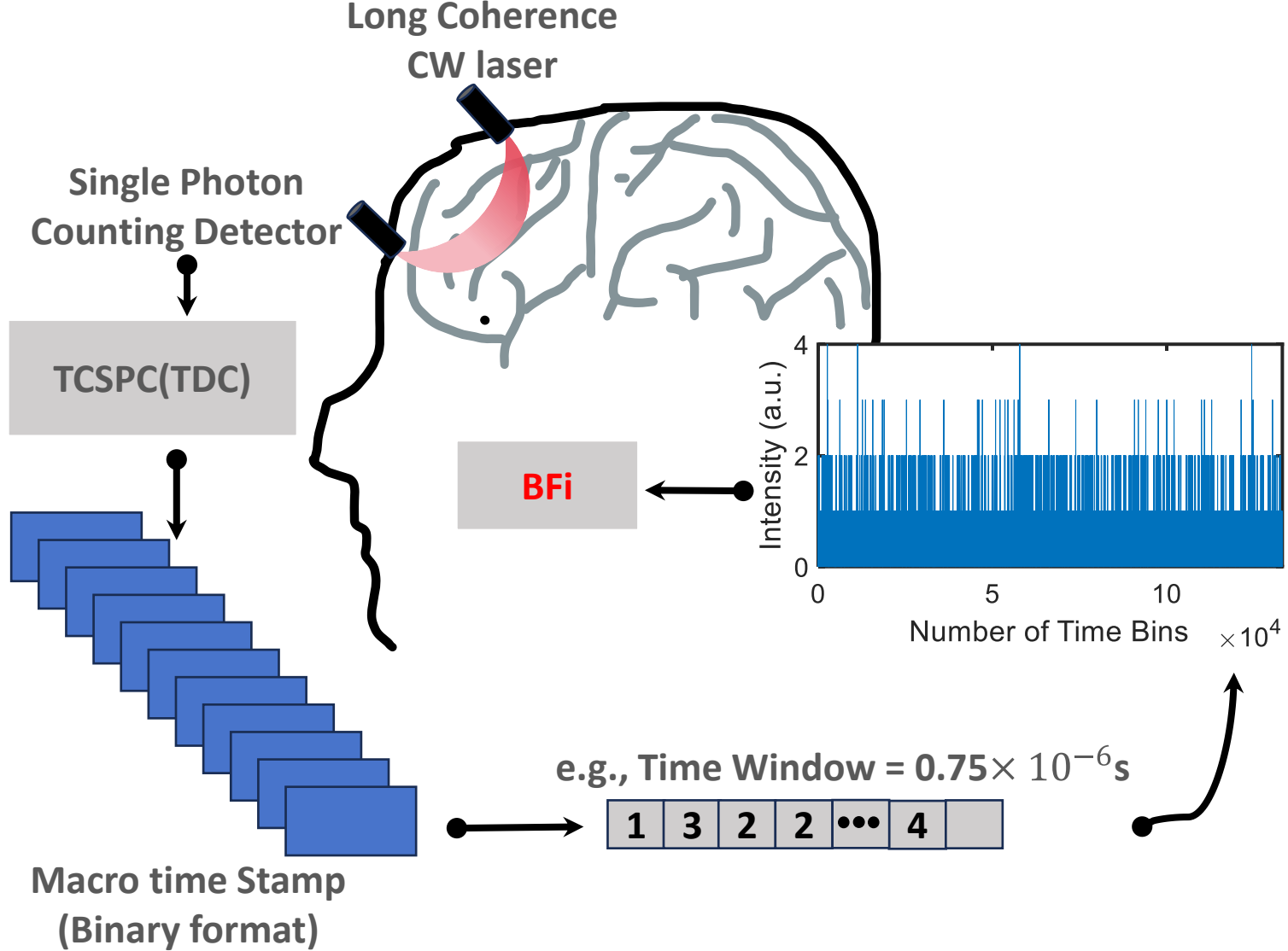

Fig. A2. Data preprocessing in conventional DCS, in which photon counts are collected using macro time stamps in a defined time window (e.g., 0.75 × $10^{-6}$ s), resulting in intensity time traces. These traces are used to calculate $g_2(\tau)$, providing information about the decorrelation speed/BFi over time. All these processes are performed offline.

## Appendix 3: Optical characterization of phantom and in vivo tissue sites

Fig. A3 and the following tables present the optical properties of the phantom and human skin (forearm and forehead) measured by FD-NIRS. Fig. A3(a) shows the absorption spectrum of the intralipid-water mixture (1:30). Although the measured absorption values were somewhat lower than literature values for pure water, particularly at a shorter wavelength where water absorption is low, they closely followed the water spectrum [58]. Fig. A3(b) displays the reduced scattering coefficients, which changed approximately linearly with the wavelength.

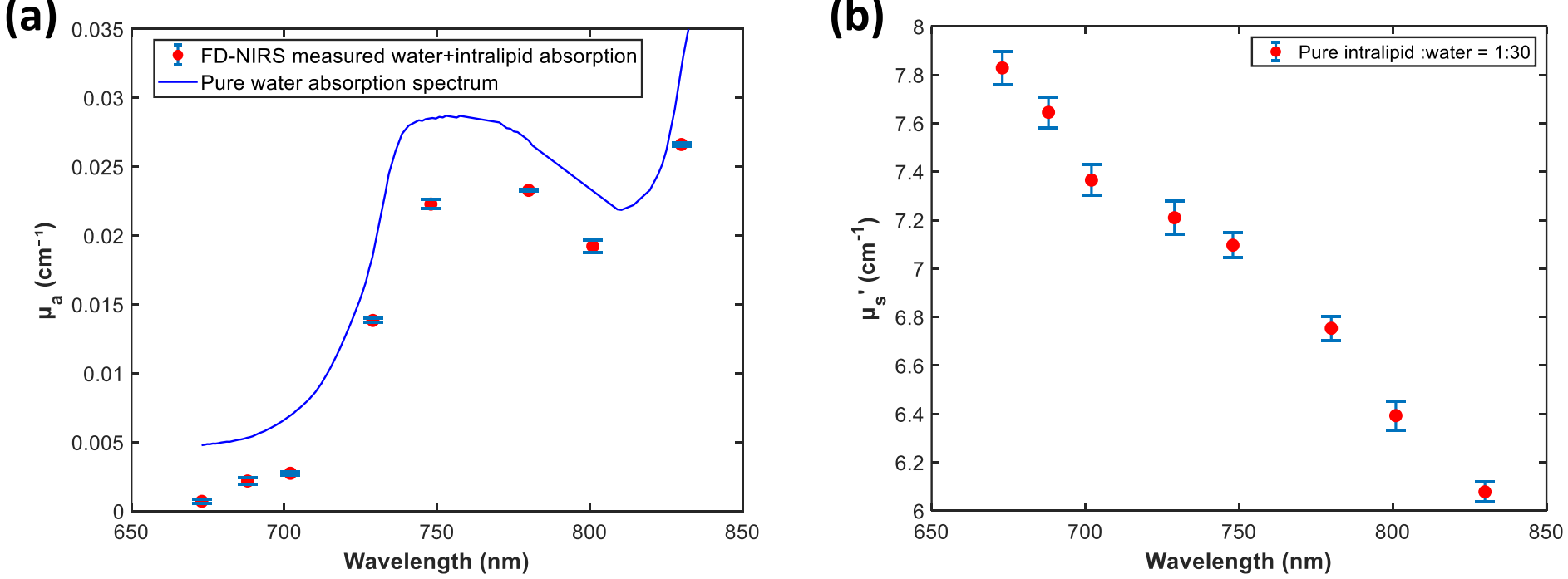

Fig. A3. (a) The absorption coefficient of the intralipid-water mixture (1:30) compared with literature values for pure water. (b) The reduced scattering coefficient of the same mixture.

Table 1 and 2 summarizes the tissue absorption, and scattering variations for the subject arm and forehead.

**Table 1. Optical parameters on the subjects' leftarm**

| | Gender | Age | $\mu_a(cm^{-1})$ | $\mu_s'(cm^{-1})$ | Average $\mu_a$ | Average $\mu_s'$ |
|---|---|---|---|---|---|---|
| Subject #1 | | | 0.12 | 4.43 | 0.12$\pm$0.021 | 4.71 $\pm$0.25 |
| | Male | 25 | 0.15 | 4.92 | | |
| | | | 0.11 | 4.77 | | |
| | | | | | | |
| | | | 0.17 | 4.07 | 0.13$\pm$0.035 | 4.23 $\pm$0.20 |
| Subject #2 | Female | 26 | 0.11 | 4.18 | | |
| | | | 0.11 | 4.45 | | |

**Table 2. Optical parameters on the subjects' forehead**

| | Gender | Age | $\mu_a(cm^{-1})$ | $\mu_s'(cm^{-1})$ | Average $\mu_a$ | Average $\mu_s'$ |
|---|---|---|---|---|---|---|
| Subject #1 | | | 0.16 | 9.91 | 0.15$\pm$0.03 | 9.98$\pm$0.11 |
| | Male | 25 | 0.11 | 10.11 | | |
| | | | 0.18 | 9.92 | | |
| | | | | | | |
| | | | 0.12 | 10.64 | 0.12$\pm$0.006 | 10.23 $\pm$0.41 |
| Subject #2 | Female | 26 | 0.12 | 9.82 | | |
| | | | 0.13 | 10.23 | | |